\documentclass[aps,prl,twocolumn,amsmath,amssymb,showpacs,superscriptaddress,notitlepage,longbibliography]{revtex4-1}
\usepackage[colorlinks=true,linkcolor=blue,anchorcolor=red,citecolor=blue,urlcolor=blue]{hyperref}
\usepackage{bm}
\usepackage{graphicx}
\usepackage{color}
\usepackage{multirow}
\usepackage{amsmath}

\begin{document}


\title{Theory for the Charge-Density-Wave Mechanism of 3D Quantum Hall Effect}

\author{Fang Qin}
\affiliation{Shenzhen Institute for Quantum Science and Engineering and Department of Physics, Southern University of Science and Technology (SUSTech), Shenzhen 518055, China}
\affiliation{CAS Key Laboratory of Quantum Information, University of Science and Technology of China, Chinese Academy of Sciences, Hefei, Anhui 230026, China}
\affiliation{Shenzhen Municipal Key-Lab for Advanced Quantum Materials and Devices, Shenzhen 518055, China}

\author{Shuai Li}
\affiliation{Shenzhen Institute for Quantum Science and Engineering and Department of Physics, Southern University of Science and Technology (SUSTech), Shenzhen 518055, China}

\author{Z. Z. Du}
\affiliation{Shenzhen Institute for Quantum Science and Engineering and Department of Physics, Southern University of Science and Technology (SUSTech), Shenzhen 518055, China}
\affiliation{Shenzhen Key Laboratory of Quantum Science and Engineering, Shenzhen 518055, China}

\author{C. M. Wang}
\affiliation{Department of Physics, Shanghai Normal University, Shanghai 200234, China}
\affiliation{Shenzhen Institute for Quantum Science and Engineering and Department of Physics, Southern University of Science and Technology (SUSTech), Shenzhen 518055, China}
\affiliation{Shenzhen Key Laboratory of Quantum Science and Engineering, Shenzhen 518055, China}

\author{Wenqing Zhang}
\affiliation{Shenzhen Institute for Quantum Science and Engineering and Department of Physics, Southern University of Science and Technology (SUSTech), Shenzhen 518055, China}
\affiliation{Shenzhen Municipal Key-Lab for Advanced Quantum Materials and Devices, Shenzhen 518055, China}

\author{Dapeng Yu}
\affiliation{Shenzhen Institute for Quantum Science and Engineering and Department of Physics, Southern University of Science and Technology (SUSTech), Shenzhen 518055, China}
\affiliation{Shenzhen Key Laboratory of Quantum Science and Engineering, Shenzhen 518055, China}

\author{Hai-Zhou Lu}
\email{Corresponding author: luhz@sustech.edu.cn}
\affiliation{Shenzhen Institute for Quantum Science and Engineering and Department of Physics, Southern University of Science and Technology (SUSTech), Shenzhen 518055, China}
\affiliation{Shenzhen Key Laboratory of Quantum Science and Engineering, Shenzhen 518055, China}

\author{X. C. Xie}
\affiliation{International Center for Quantum Materials, School of Physics, Peking University, Beijing 100871, China}
\affiliation{CAS Center for Excellence in Topological Quantum Computation, University of Chinese Academy of Sciences, Beijing 100190, China}
\affiliation{Beijing Academy of Quantum Information Sciences, West Building 3, No. 10, Xibeiwang East Road, Haidian District, Beijing 100193, China}

\date{\today }

\begin{abstract}
The charge-density-wave (CDW) mechanism of the 3D quantum Hall effect has been observed recently in ZrTe$_5$ [\href{http://dx.doi.org/10.1038/s41586-019-1180-9}{Tang \emph{et al}., Nature \textbf{569}, 537 (2019)}].
Different from previous cases, the CDW forms on a one-dimensional (1D) band of Landau levels, which strongly depends on the magnetic field. However, its theory is still lacking. We develop a theory for the CDW mechanism of 3D quantum Hall effect. The theory can capture the main features in the experiments. We find a magnetic field induced second-order phase transition to the CDW phase. We find that electron-phonon interactions, rather than electron-electron interactions, dominate the order parameter. We extract the electron-phonon coupling constant from the non-Ohmic $I$-$V$ relation.
We point out a commensurate-incommensurate CDW crossover in the experiment. More importantly, our theory explores a rare case, in which a magnetic field can induce an order-parameter phase transition in one direction but a topological phase transition in other two  directions, both depend on one magnetic field.
\end{abstract}

\maketitle

\begin{figure}[htpb]
\centering
\includegraphics[width=0.48\textwidth]{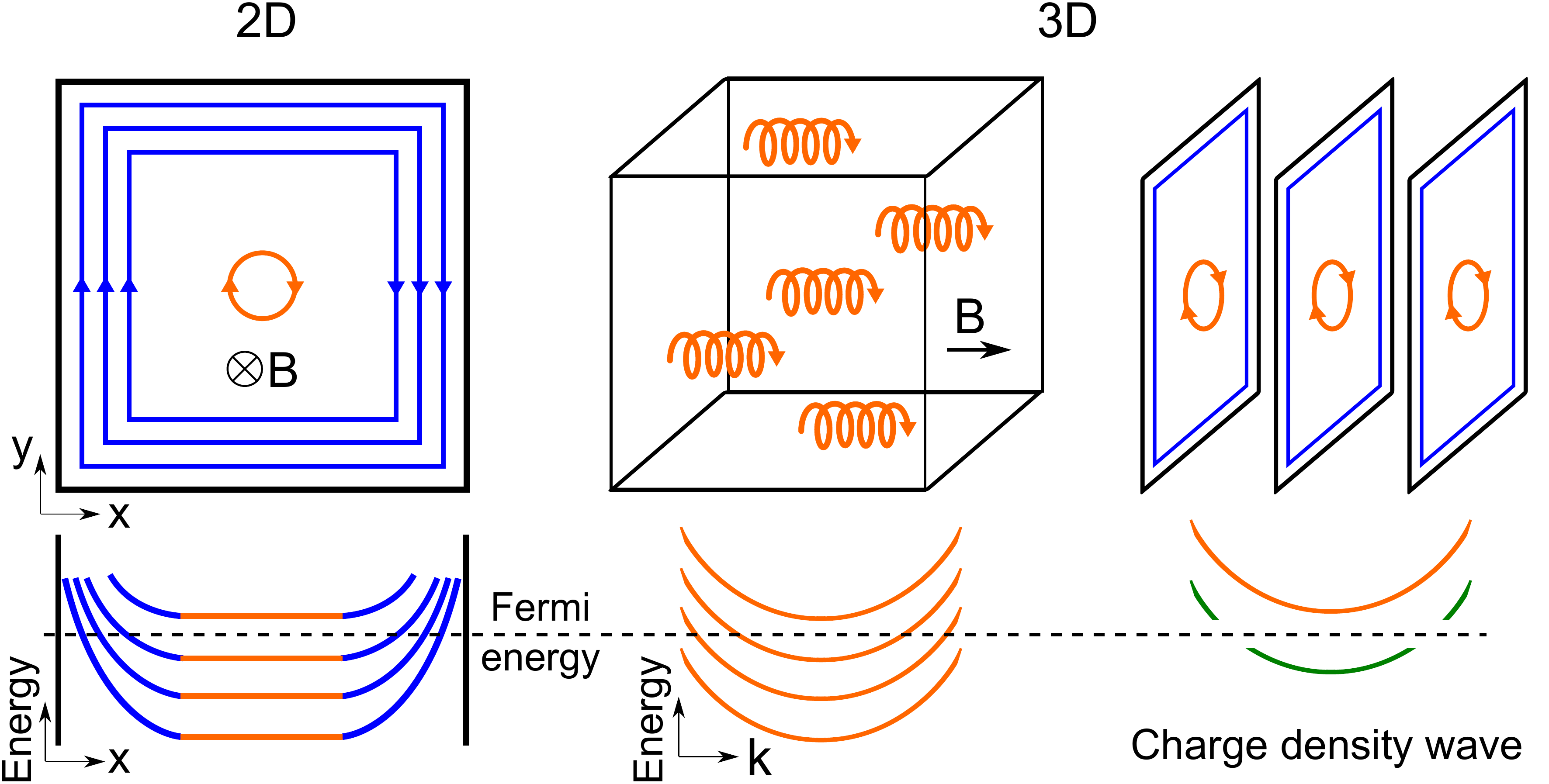}
\caption{Left: In 2D, the quantum Hall effect arises when only the edge states (blue) conduct electrons, while the interior bulk states are insulating as the Fermi energy lies between the Landau levels. Center: In 3D, the Landau levels turn to 1D bands of Landau levels that disperse with the momentum ($k$) along the direction of magnetic field.
The quantum Hall effect is difficult in 3D because the bulk is metallic as the Fermi energy always crosses some Landau bands.
Right: The charge density wave may gap the Landau band, so that the bulk is insulating and the quantum Hall effect can be observed.} \label{Fig:2D3D}
\end{figure}

{\color{blue}\emph{Introduction.}} --
The quantum Hall effect is one of the most important discoveries in physics \cite{Klitzing80prl,Tsui82prl,Laughlin83prl,Thouless82prl}. It arises from the Landau levels of two-dimensional (2D) electron gas in a strong magnetic field (Fig.~\ref{Fig:2D3D} Left). When the Fermi energy lies between two Landau levels, the interior of the electron gas is insulating but the deformed Landau levels at the edges can transport electrons dissipationlessly, leading to the quantized Hall resistance and vanishing longitudinal resistance of the quantum Hall effect.
The quantum Hall effect is difficult in 3D, where the Landau levels turn to a series of 1D bands of Landau level dispersing with the momentum along the direction of magnetic field (Fig.~\ref{Fig:2D3D} Center). Because the Fermi energy always crosses some Landau bands, the interior is metallic, which buries the quantization of the edge states, so the quantum Hall effect is usually observed in 2D systems~\cite{ZhangYB05nat}.
Nevertheless, searching for a 3D quantum Hall effect has been lasting for more than 30 years \cite{Lu18nsr,Halperin87jjap,Montambaux90prb,Kohmoto92prb,Koshino01prl,Bernevig07prl,Stormer86prl,Cooper89prl,Hannahs89prl,Hill98prb,Cao12prl,Masuda16sa,Liu16nc,WangCM17prl, ZhangC17nc,Uchida17nc,Schumann18prl,ZhangC19nat,LiuJY19arXiv,LiH20prl,ChengSG20prb,Wang20prb,Ma20arxiv}.
One of the famous proposals for the 3D quantum Hall effect relies on the charge density wave (CDW), which may gap the 1D Landau band so that the bulk is insulating. In real space, the CDW splits the 3D electron gas into decoupled 2D quantum Hall layers to realize a 3D quantum Hall effect (Fig.~\ref{Fig:2D3D} Right) \cite{Halperin87jjap}. Quite different from the known cases \cite{Gruner00book,Giamarch04book,Lee79prb}, the CDW of Landau bands depends on the magnetic field strongly \cite{Celli65pr,Yakovenko93prb,Moller07prl,ZhangXT17prb,Yang11prb,Song17prb,Trescher17prbrc,Trescher18prb,Pan19prb}. 
Recently, the CDW mechanism of the 3D quantum Hall effect has been observed in 3D crystals of ZrTe$_5$~\cite{Tang19nat}, providing a platform to study this rare phase of matter where both order parameter and topological number coexist.

In this Letter, we develop a theory for the CDW mechanism of 3D quantum Hall effect. The theory captures the main features in the experiment of ZrTe$_5$ at the quantitative level.
We find that electron-phonon interactions dominate the formation of the CDW, instead of electron-electron interactions. We extract electron-phonon coupling constant from the non-Ohmic $I$-$V$ relation. We point out a crossover between commensurate and incommensurate CDWs, tunable by the magnetic field. More importantly, the theory addresses a rare but experiment-accessible scenario, described by an order parameter along one direction but a topological Chern number in other two directions, both tunable by one magnetic field.

\begin{figure}[tpb]
\centering
\includegraphics[width=0.48\textwidth]{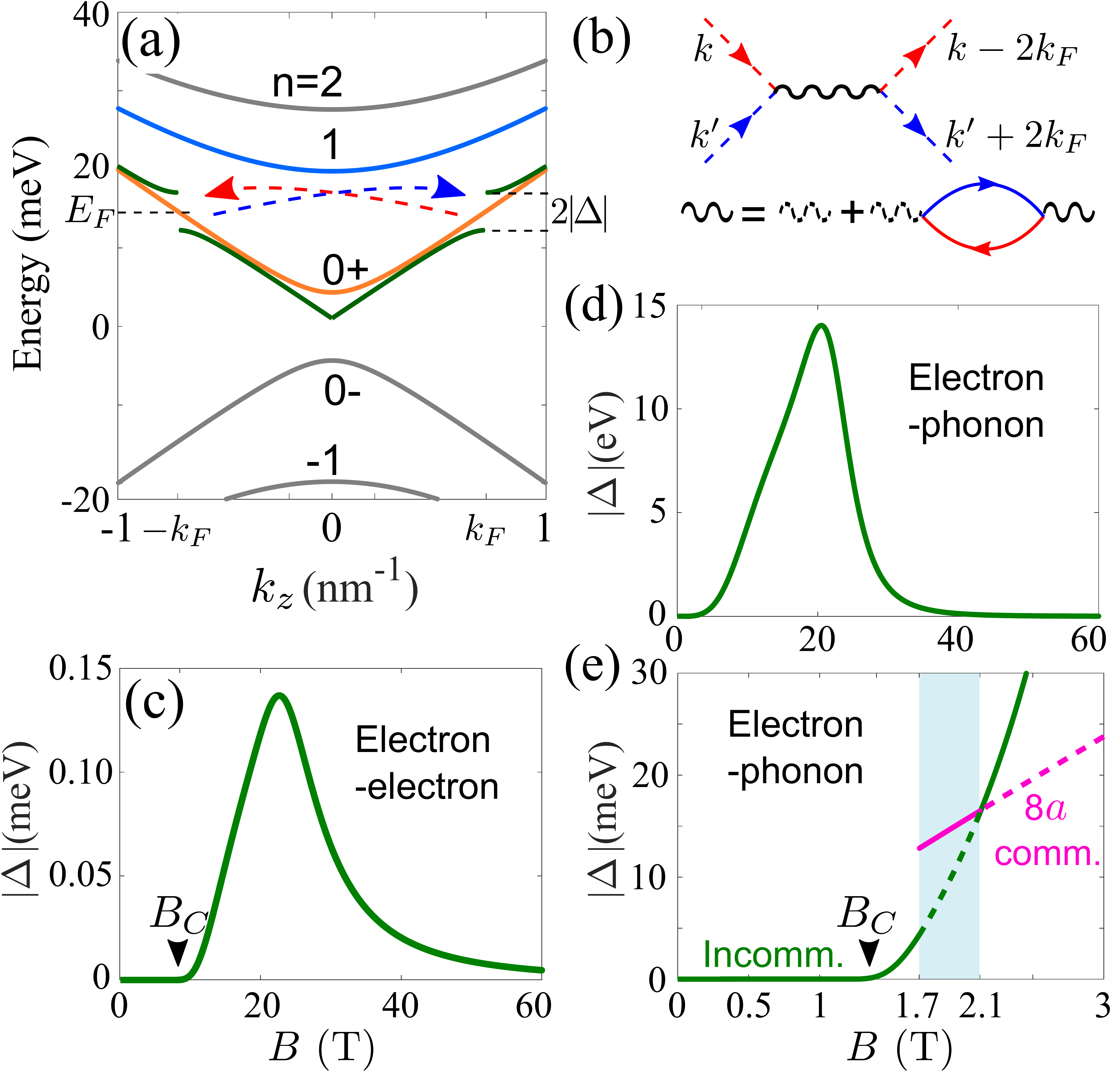}
\caption{(a) The 1D energy bands of Landau levels dispersing with the $z$-direction wave vector $k_z$ in a $z$-direction magnetic field $B=1.6$ T. The CDW opens the gap ($2|\Delta|$) at the Fermi energy $E_F$. $n$ marks the indices of the Landau bands. $n=0\pm$ are the lowest Landau bands. (b) Up: The $g$-ology, which is a diagrammatic representation of the interaction and scattering processes (arrowed $\pm k_F$) involved in the charge density wave and competing phases~\cite{Giamarch04book}. Down: The diagrams for the Yukawa potential. The solid wavy line stands for interactions under the random phase approximation~\cite{Bruus04book}, the dashed wavy line represents the bare Coulomb interaction, and the solid line represents the bare electronic propagator. [(c)-(e)] The calculated CDW order parameters for electron-electron (c) and electron-phonon [(d),(e)] interactions, respectively. $B_C$ indicates a threshold magnetic field at which there is a second-order phase transition as $\Delta$ overcomes temperature. ``Incomm." and ``8$a$ comm." indicate that incommensurate and commensurate (CDW wavelength/lattice constant =8) CDWs are assumed, respectively. The parameters are $v_{x}=9\times10^5 $ m/s, $v_{y}=1.9\times10^5 $ m/s, $v_{z}=0.3\times10^5$ m/s, $M_{0}=-4.7$ meV, $M_{\perp}=150$ meV$\cdot$nm$^{2}$, $M_{z}=0.01M_{\perp}$, $a=7.25$ \AA~\cite{Tang19nat,Jiang17prbrc}, $n_{0}=8.87\times10^{16}$ cm$^{-3}$,
$\epsilon_{r}=25.3$ \cite{Hochberg18prd}, and the electron-phonon coupling constant $g_{0}=537.3$ eV$\cdot$nm$^{-1}$ (determined by comparing with the nonlinear $I$-$V$ data \cite{Tang19nat} in Fig.~\ref{fig:non-Ohmic} (h)), and $T=0$ K. } \label{Fig:CDW}
\end{figure}

{\color{blue}\emph{1D Landau band in the quantum limit.}} --
We start with a generic Dirac model ~\cite{ChenRY15prl}
\begin{eqnarray}
\hat{{\cal H}}({\bf k}) &=& \hbar(v_{x}k_{x}\tau_{x}\otimes\sigma_{z} + v_{y}k_{y}\tau_{y}\otimes\sigma_{0} + v_{z} k_{z}\tau_{x}\otimes\sigma_{x}) \nonumber\\
&&+ [M_{0} + M_{1}(v_{x}^{2}k_{x}^{2} + v_{y}^{2}k_{y}^{2}) + M_z k_{z}^{2}]\tau_{z}\otimes\sigma_{0},
\end{eqnarray}
where $k_{x/y}=-i\partial_{x/y}$, $\tau_{x,y,z,0}$, and $\sigma_{x,y,z,0}$ are Pauli matrices and unit matrix for orbital and spin degrees of freedom, and $M_{0,1,z}$, $v_{x,y,z}$ are the model parameters. This model can describe various semimetals and insulators~\cite{Weng14prx,Fan17srep,Manzoni16prl,ChenRY15prb,Liu16nc,Nair18prbrc,Li16np,LiXB16prl,ChenZG17pnas,Jiang17prbrc,Xiong17prb,Xu18prl}.
A uniform $z$-direction (crystal $b$ direction) magnetic field ${\bf B}=(0,0,B)$ is considered by the Landau gauge vector potential ${\bf A}=(-By,0,0)$, which shifts $k_x$ to $k_x-eBy/\hbar$, where
$-e$ is the electron charge and $\hbar$ is the reduced Planck's constant.
The magnetic field splits the energy spectrum into a series of 1D bands of Landau levels, dispersing with $k_z$ [Fig.~\ref{Fig:CDW} (a)].

We will focus on the quantum limit, in which the Fermi energy $E_F$ crosses only the $n=0+$ Landau band \cite{ZhengGL16prb}.
At the critical magnetic field $B_{\mathrm{Q}}$ when entering the quantum limit, $E_F=E^{(0+)}_{k_z=k_F}=E^{(1)}_{k_z=0}$,
where the Fermi wave vector \cite{Lu15prb-QL}
\begin{eqnarray}\label{Eq:kF-n0}
k_F = 2\pi^{2}\hbar n_{0}/eB,
\end{eqnarray}
$n_0$ is carrier density, the energy dispersion of the $n=0+$ Landau band $E^{(0+)}_{k_z}$ = $\sqrt{(\hbar v_{z}k_{z})^{2} + (M_{0} + M_{\perp}/\ell_{B}^{2} + M_z  k_z^2)^2 }$,
$M_{\perp}=M_{1}v_{x}v_{y}$, the magnetic length is $\ell_{B}=\sqrt{\hbar/eB}$,
the bottom of the $n=1$ Landau band $E^{(1)}_{k_z=0}$ = $\sqrt{ (M_{0} + 3 M_{\perp}/\ell_{B}^{2} )^{2} +  2 v_xv_y \hbar^{2}/\ell_{B}^{2} }$.
Using $B_{\mathrm{Q}}=1.3$ T in the above equations, $n_{0}$ is found as $8.87\times10^{16}$ cm$^{-3}$, comparable with the experiment \cite{Tang19nat}, showing that our model and parameters can capture the noninteracting energy spectrum. At this low carrier density, the pocket at the $M$ point does not contribute~\cite{Tang19nat,ZhangJL19prl}.

{\color{blue}\emph{Theory of CDW for the Landau band.}} --
We study the CDW of the $0+$ Landau band by using a mean-field approach, which can capture the physics of 1D CDWs \cite{Gruner00book, Giamarch04book}. Different from previous theories (e.g., \cite{Lee79prb}), the 1D Landau band here strongly depends on the magnetic field, e.g., the changing $k_F$ in Eq.~\eqref{Eq:kF-n0}, the nesting momentum $k_{cdw}$, and CDW wavelength $\lambda_{cdw}$.

As shown by the $g$-ology diagram in Fig.~\ref{Fig:CDW}~(b), the CDW gap (described by the order parameter $\Delta$) can be opened by the coupling between the electrons near $k_F$ and $-k_F$, through either electron-electron or electron-phonon interactions along the $z$ direction.
The electron-electron interaction reads \cite{supp,Bruus04book,Pan19prb,Song17prb,Jishi13book}
\begin{align}
\hat{H}_{ee}
&= -\sum_{{\bf k}}|\Delta|\left(e^{i\phi}\hat{d}^{\dag}_{{\bf k}+}\hat{d}_{{\bf k}-} + h.c. \right) + \frac{2|\Delta|^{2}V}{U(2k_F)},
\end{align}
where the order parameter is defined as $\Delta= \Delta_{ee}=[U(2k_F)/2V]\sum_{{\bf k}}\langle \hat{d}^{\dag}_{{\bf k}-2k_{F}{\bf e}_{z}}\hat{d}_{{\bf k}} \rangle$ and $V$ is the volume. $\Delta=|\Delta|e^{i\phi}$, where $\phi$ is the phase. $\hat{d}^{\dag}_{{\bf k}\pm}$ and $\hat{d}_{{\bf k}\pm}$ are the creation and annihilation operators in the vicinity of $\mp k_F$, respectively, where ${\bf k}\pm\equiv k_z\pm k_F$. As shown in Fig.~\ref{Fig:CDW} (b), the electron-electron potential takes the Yukawa form~\cite{Abrikosov98prb}
$U(2k_F) = e^{2}/\{\epsilon_{r}\epsilon_{0}[(2k_{F})^{2} + \kappa^{2}]\}$, where $\epsilon_{r}$ ($\epsilon_{0}$) is the relative (vacuum) dielectric constant and $1/\kappa$ is the screening length. Under the random phase approximation [Fig.~\ref{Fig:CDW}~(b)], we have $\kappa=\sqrt{e^{3}B/(4\pi^{2}\epsilon\hbar^{2}v_{F})}$ (Eq.(S18) in \cite{supp}) with $\epsilon=\epsilon_{0}\epsilon_{r}$.
The Hamiltonians for electron-phonon interaction and phonons can be, respectively, written as~\cite{Bruus04book,Gruner00book,Roy14prb,Gruner88rmp}
\begin{eqnarray}
\hat{H}_{\mathrm{e-ph}}
= \sum_{{\bf k}}|\Delta|(e^{i\phi}\hat{d}^{\dag}_{{\bf k}+}\hat{d}_{{\bf k}-} + h.c. ),~\hat{H}_{\rm ph} = \sum_{{\bf q}}\hbar\omega_{{\bf q}}\hat{b}^{\dag}_{{\bf q}}\hat{b}_{{\bf q}}, \nonumber\\
\end{eqnarray}
where $\Delta= \Delta_{e-ph}= (\alpha_{\bf q}/V) (\langle \hat{b}_{\bf q} \rangle + \langle \hat{b}^{\dag}_{-{\bf q}} \rangle)$, $\hat{b}^{\dag}_{{\bf q}}$ and $\hat{b}_{{\bf q}}$ are the creation and annihilation operators for the phonons with momentum ${\bf q}=\pm2k_{F}{\bf e}_{z}$, the electron-phonon coupling~\cite{Bruus04book}
$\alpha_{\bf q}$ also takes the Yukawa form (Sec. SIV(B) of \cite{supp}).
Near $\pm k_F$, the mean-field Hamiltonian of the $0+$ Landau band can be written as (Sec. SIV of \cite{supp})
\begin{align}\label{Hn0}
{\cal H}^{0+}_{k_z} =\left(
  \begin{array}{cc}
    \hbar v_F (k_z \pm k_F) & \Delta \\
    \Delta^{*} & -\hbar v_F (k_z \pm k_F) \\
  \end{array}
\right),
\end{align}
where $\hbar v_F\equiv | \partial E^{(0+)}_{k_z}/\partial k_z |_{k_z=k_F} $ (Sec. SIII of \cite{supp}). The eigen energies of ${\cal H}^{0+}_{k_z} $ can be found as
$E_{k_z}$ = $E_F \pm \text{sgn}(k_{z}\mp k_F)$ $\sqrt{[v_F\hbar(k_{z}\mp k_F)]^{2} + |\Delta|^{2} } $ near $\pm k_F$ [green curves in Fig.~\ref{Fig:CDW}~(a)], respectively,
where $\text{sgn}(x)$ is the sign function.

The CDW order parameter is calculated self-consistently from the gap equation defined by $\partial E_{g}/\partial|\Delta|=0$, where the ground-state energy
$E_{g} \equiv \langle\hat{H}_{m}\rangle $ is found as (Sec. SV of \cite{supp})
\begin{eqnarray}\label{eq:omegaT}
E_{g}
= \sum_{{\bf k}} (E_{k_z} - E_{F}) \Theta\left(E_{F} - E_{k_z}\right) + \frac{|\Delta|^{2}V}{g_{2k_F}},
\end{eqnarray}
where $E_g$ includes the phonon part,
$\Theta(x)$ is the step function, $\hat{H}_{m} = \sum_{\bf k} \hat{\Psi}^{\dag}_{\bf k} {\cal H}^{0+}_{k_z} \hat{\Psi}_{\bf k} + |\Delta|^{2}V/g_{2k_F}$,
$\hat{\Psi}_{\bf k}\equiv (\hat{d}_{\mathbf{k}+},\hat{d}_{\mathbf{k}-})^T$, and ${\cal H}^{0+}_{k_z}$ has been given in Eq. (\ref{Hn0}). The coupling $g_{2k_F}=e^{2}/\{2\epsilon[(2k_{F})^{2} + \kappa^{2}]\}$ for electron-electron interactions and $g_{2k_F}=g_{0}/[(2k_F)^{2} + \kappa^{2}]^{2}$ for electron-phonon interactions with the coupling constant $g_{0}$ (Sec. SIV(B) of \cite{supp}). The second positive term is from the mean-field phonon Hamiltonian (Eq. (S28) in \cite{supp}). As a function of the order parameter, Eq. (6) reduces to a minimum value (GS energy) at a finite gap as shown in Fig. S3 of \cite{supp}. Different from no-magnetic-field theories, here the summation $\sum_{k_x,k_y} =  S_{xy}/(2\pi\ell_{B}^{2})$ gives the Landau degeneracy, with the area $S_{xy}$ in the $x-y$ plane, $V=S_{xy}L_{z}$, and the length $L_{z}$ along the $z$ direction.

\begin{figure}[tbp]
\centering
\includegraphics[width=0.49\textwidth]{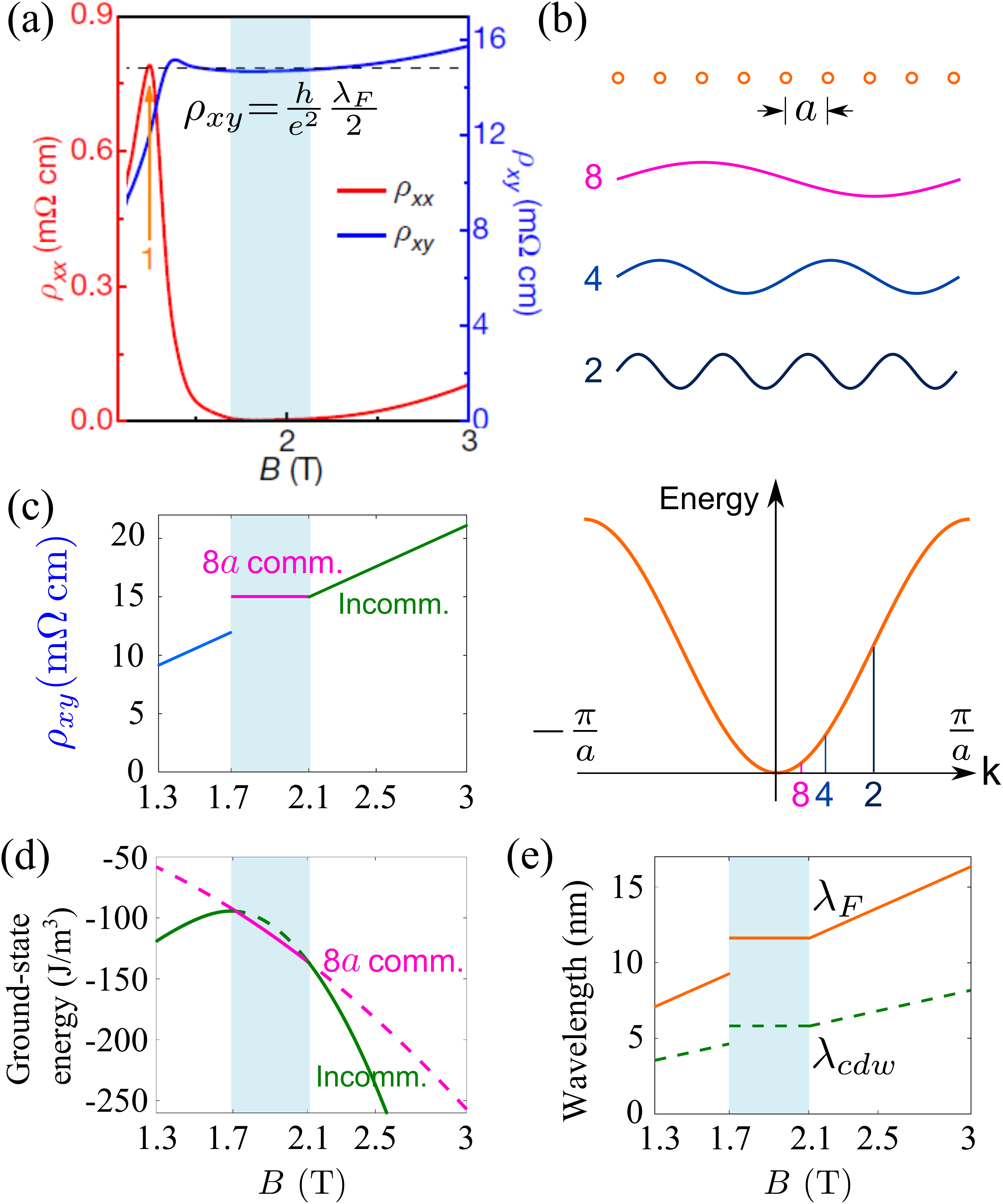}
\caption{(a) The Hall ($\rho_{xy}$) and longitudinal ($\rho_{xx}$) resistivities adapted from the experiment \cite{Tang19nat}.
(b) Schematic of the commensurate CDWs, whose wavelengths are integer times of the lattice constant $a$.
(c) Our understanding to $\rho_{xy}$. $B\in[1.3,1.7]$ T, $\rho_{xy}$ is not quantized due to the broadening of the $n=1$ Landau band bottom [see also Fig.~\ref{fig:non-Ohmic}~(e)]; $B\in[1.7,2.1]$ T, a commensurate CDW pins $\lambda_{cdw}$ and $\lambda_F$, leading to the plateau of $\rho_{xy}$; $B\in[2.1,3]$ T,
the incommensurate CDW takes over, so $\rho_{xy}\propto B$.
(d) Ground-state energies $E_g$ (per unit volume) of incommensurate and commensurate ($\lambda_{cdw}/a=8$) CDWs, which shows that the commensurate (incommensurate) CDW has lower energy when $B\in [1.7,2.1]$ ($[2.1,3]$ T).
(e) The Fermi ($\lambda_F$) and CDW ($\lambda_{cdw}$) wavelengthes. }\label{fig:DeltaBT0}
\end{figure}

At extremely low temperatures, i.e., $T\rightarrow0$, the finite-temperature gap equation can be expressed as (Sec. SVI(C) of \cite{supp})
\begin{align}\label{eq:gap-EqT}
\int_{0}^{v_{F}\hbar k_{F}} \frac{1}{1 + e^{-E_{T}(t,\Delta)/k_{B}T}} \frac{dt}{ E_{T}(t,\Delta) }
 = \frac{4\pi^{2}\hbar^{2}v_{F}}{g_{2k_F}eB},
\end{align} where $E_{T}(t,\Delta)= \sqrt{t^{2} + |\Delta|^{2} }$, $k_B$ is the Boltzmann constant, and $T$ is the temperature.
We use the Ginzburg criterion~\cite{Gruner00book,Levanyuk59jetp,Ginzburg60jetp,Coleman15book,Lee73prl} to justify the mean-field approximation at the experimental finite temperatures (Sec. SIV of \cite{supp}).
Also, we find that the commensurability energy from the ionic potential of the crystal \cite{Lee74ssc,Kagoshima88book,Gruner00book} can be ignored (Sec. SVII of \cite{supp}).

\begin{figure}[th]
\centering
\includegraphics[width=0.45\textwidth]{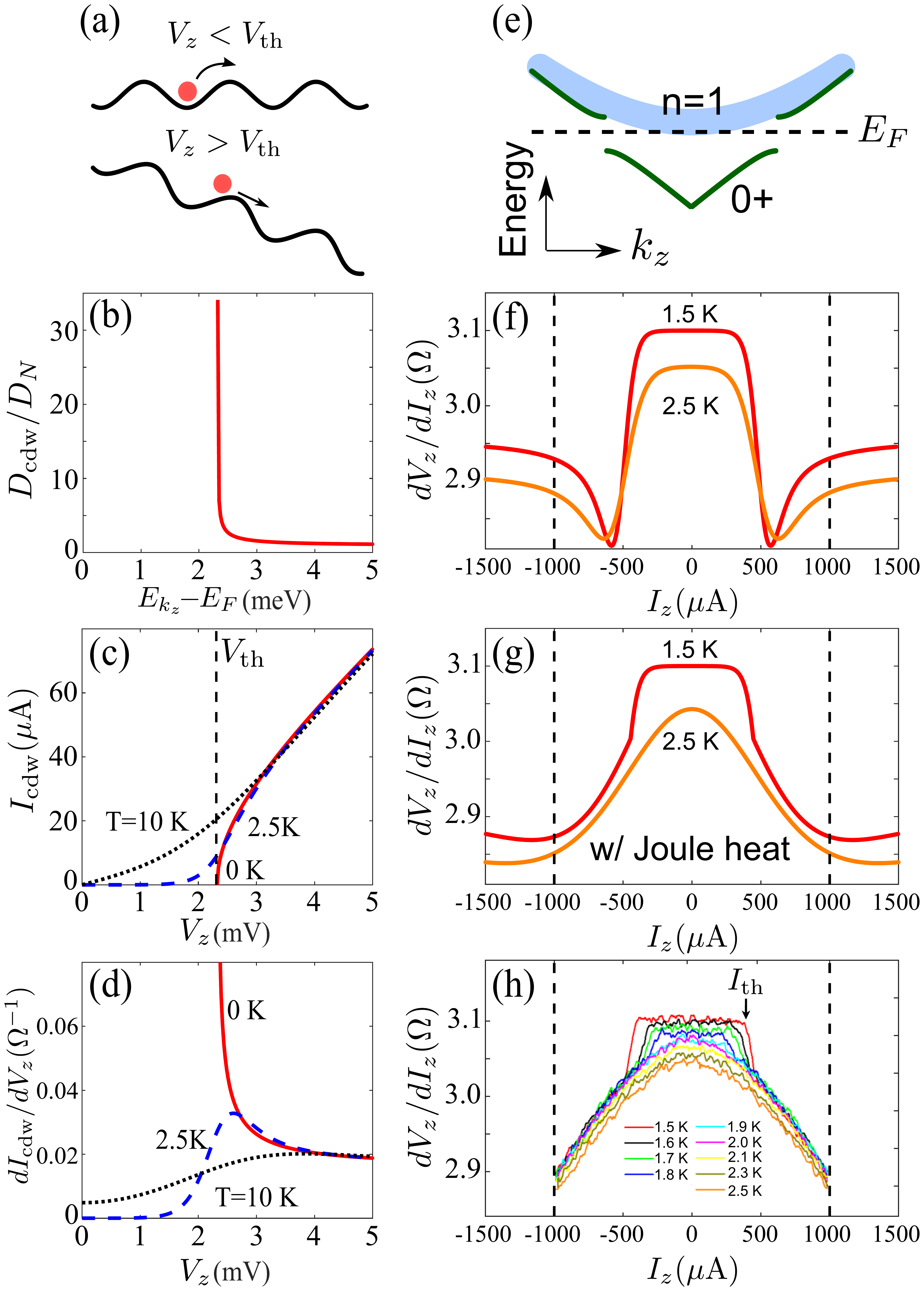}
\caption{(a) A bias voltage $V_{z}$ has to overcome the threshold voltage $V_{\mathrm{th}}$ of CDW to yield a current, leading to the non-Ohmic $I$-$V$ relation.
[(b)-(d)] At $B=1.6 $ T, CDW density of states (b), non-Ohmic relation between the tunneling current $I_{cdw}$ and $V_{z}$ (c), differential conductance $dI_{cdw}/dV_z$ (d).
(e) At $B=1.55 $ T, the Fermi energy $E_F$ is assumed to cross both the CDW-gapped $n=0+$ and broadened $n=1$ Landau bands.
[(f) and (g)] Differential resistance $dV_z/dI_z$ as a function of the $z$-direction current at $B=1.55 $ T and different temperatures, without (f) and assuming the Joule heat (g). The parameters $\alpha_{1}=7$, $G_{cdw}^{(1)}=322.58$ m$\Omega^{-1}$, $G_{N}^{(1)}=16.64$ m$\Omega^{-1}$, $G_{cdw}^{(2)}=293.25$ m$\Omega^{-1}$, and $G_{N}^{(2)}=49.75$ m$\Omega^{-1}$ at $T=1.5 $ K; $\alpha_{2}=5$, $G_{cdw}^{(3)}=294.12$ m$\Omega^{-1}$, and $G_{N}^{(3)}=53.76$ m$\Omega^{-1}$ at $T=2.5 $ K.
(h) Experimental data of $dV_z/dI_z$~\cite{Tang19nat}.
}\label{fig:non-Ohmic}
\end{figure}

{\color{blue}\emph{Electron-electron or electron-phonon interactions?}}--
As shown in Fig.~\ref{Fig:CDW}~(c), the order parameter calculated using electron-electron interactions is sizable only beyond a threshold magnetic $B_C$ about 10 T, an order larger than
those in the experiments [Fig.~\ref{fig:DeltaBT0}~(a)]. On the other hand, for electron-phonon interactions with a proper coupling constant ($g_{0}=537.3$ eV$\cdot$nm$^{-1}$, determined by the non-Ohmic I-V relation [Fig. \ref{fig:non-Ohmic} (h)]), the threshold $B_C$ could be less than 1.5 T and $\Delta$ could be of several to tens of meV [Fig.~\ref{Fig:CDW}~(e)], both consistent with the experiment. Therefore, electron-phonon interactions may be the mechanism in the ZrTe$_5$ experiment.

{\color{blue}\emph{Commensurate-incommensurate crossover.}}
In the experiment, the plateau of the Hall resistivity covers a wide range from 1.7 to 2.1 T, which is surprising for the following reason.
According to Fig.~\ref{Fig:2D3D}, the Hall conductivity in units of $e^2/h$ is given by the number of the CDW layers $\sigma_{xy}=\frac{e^2}{h}/\lambda_{cdw}$ per unit length, where $\lambda_{cdw}$ is the CDW wavelength, so the height of plateau should be $\rho_{xy}=1/\sigma_{xy}=\frac{h}{e^2} \lambda_{cdw}$ when $\sigma_{xx}=0$. It is known that the CDW wavelength $\lambda_{cdw}$ is related to the Fermi wavelength as \cite{Gruner00book} (Sec. SV of \cite{supp})
\begin{eqnarray}
\lambda_{cdw} = \lambda_{F}/2=\pi/k_F.
\end{eqnarray}
According to Eq.~(\ref{Eq:kF-n0}), $k_F$ decreases with the magnetic field, leading to a $\lambda_{cdw}$ linearly increasing with the magnetic field [e.g., $B>2.1$ T in Fig.~\ref{fig:DeltaBT0}~(e)], so $\rho_{xy}$ should increase linearly with $B$. That is why the plateau in Fig.~\ref{fig:DeltaBT0}~(a) is surprising.

The observed $\rho_{xy}$ plateau between 1.7 and 2.1 T implies that there is a commensurate CDW, i.e., the CDW wavelength is pinned at integer times of the lattice constant $a$ [Fig.~\ref{fig:DeltaBT0}~(b)]. According to the experiment, $\lambda_{cdw}/a=8.1\pm 0.8$ \cite{Tang19nat}.
We compare the ground-state energies of commensurate ($\lambda_{cdw}/a=8$) and incommensurate CDWs near 2.1 T, which can be obtained by minimizing the ground-state energy $E_g $ in Eq.~(\ref{eq:omegaT}). As shown in Fig.~\ref{fig:DeltaBT0}~(d), the commensurate (incommensurate) CDW has lower energy for $B\in [1.7,2.1]$ ([2.1, 3]) T, so there is a crossover between the commensurate and incommensurate CDWs [$B=2.1$ T in Fig.~\ref{fig:DeltaBT0}~(c)]. In the range $B\in [1.7,2.1]$ T, the fixed $\lambda_{cdw}$ means a fixed Fermi energy, i.e, the system is a grand canonical ensemble and the number of carriers can change. By contrast, the number of carriers in the incommensurate CDW phase cannot change. Therefore, the change of electrons leads to lower ground-state energy of the commensurate CDW phase in the range $B\in [1.7,2.1]$ T. Further increasing the magnetic field above 2.1 T, the magnetic field will push the Fermi energy lower (eventually to the band bottom), so there is a crossover from commensurate to incommensurate CDW phase as a function of the magnetic field. These are unique properties of this magnetic field-induced CDW.

{\color{blue}\emph{Non-Ohmic $I$-$V$ relation.}}--An evidence of CDW is the non-Ohmic $I$-$V$ relation~\cite{Bardeen79prl,Bardeen80prl}, because a bias voltage has to overcome the barriers of CDW [Fig.~\ref{fig:non-Ohmic}~(a)], which can be used to determine $\Delta$ and more importantly the electron-phonon interaction coupling constant $g_{0}$ by comparing with our theory. The tunneling current $I_{cdw}$ is found as~\cite{Mahan00book,Gruner00book}
\begin{align}
I_{cdw}\! =\! \frac{e}{h}|\mathcal{T}|^2\!\int_{\!-\!\infty}^{\infty}\!d\epsilon D_{cdw}(\epsilon)D_{N}(\epsilon\!+\!eV_{z})[f(\epsilon)\! - \!f(\epsilon\!+\!eV_{z})],
\end{align}
with the density of states [Fig.~\ref{fig:non-Ohmic}~(b)]
$D_{cdw}(E_{k_z})/D_{N}(0)$ =
$|E_{k_z} - E_F|\Theta(|E_{k_z} - E_F|-|\Delta|)/\sqrt{ (E_{k_z} - E_F)^{2} - |\Delta|^{2} }$ (Sec. SVIII(A) of \cite{supp}),
where the normal ($N$) density of states $D_{N}(0)$ and tunneling matrix element $\mathcal{T}$ are assumed energy-independent, and $f(x)=1/[1 + e^{x/(k_{B}T)}]$ is the Fermi function~\cite{Pathria96book}.
Figure~\ref{fig:non-Ohmic}~(c) shows the non-Ohmic $I_{cdw}$-$V_{z}$ relation at different temperatures. At zero temperature, there is no tunneling current below the threshold voltage $V_{th}\equiv|\Delta|/e$. Finite temperatures can lead to a small tunneling current for $|V_z|<V_{th}$.
Figure~\ref{fig:non-Ohmic}~(d) shows the differential conductance $dI_{cdw}/dV_{z} $ as a function of $V_z$ at different temperatures, where the peak near the threshold $V_{th}$ at $T=2.5 $ K is due to the abrupt increase of $I_{cdw}$ across the threshold and is smeared at higher temperatures.

Figure~\ref{fig:non-Ohmic}~(h) shows the differential resistance $dV_z/dI_z$ in the experiment \cite{Tang19nat}. There is a plateau below the threshold current $I_{th}\approx 450~\mu$ A, besides the non-Ohmic behavior above $I_{th}$. This implies that besides the $0+$ Landau band, there is another Ohmic channel, likely the broadened $+1$ band bottom which lasts till $B=1.7$ T [Fig.~\ref{fig:non-Ohmic}~(e)]. Therefore, we model the current as $I_{z} = I_{cdw} + I_{N}$~\cite{Gruner88rmp,Gruner85pr}, where $I_{cdw}$ is the CDW current from the $0+$ band and the normal band is assumed to satisfy the Ohmic law $I_{N}=G_{N}V_{z}$.
We reproduce the Ohmic plateau and non-Ohmic $I_{z}$-$V_{z}$ relation at different temperatures [Fig.~\ref{fig:non-Ohmic}~(g)].
Using $I_{th}$ in the experiment, we find that $g_{0}=537.3$ eV$\cdot$nm$^{-1}$.
For $T$=1.5 K, we assume that
$I_{z}=
I_{cdw}^{(1)}(T) + I_{N}^{(1)}$ for $I_{z}<I_{th}$ and
$I_z=I_{cdw}^{(2)}(\alpha_{1} T) + I_{N}^{(2)}$ for $I_{z}>I_{th}$; for $T=2.5 $ K, $I_{z}=I_{cdw}^{(3)}(\alpha_{2} T) + I_{N}^{(3)}$, where $\alpha_{1,2}$ describe the Joule heat from the abrupt current increase. Without the Joule heat, $dV_z/dI_z$ shows a dip near $I_{th}$ [Fig.~\ref{fig:non-Ohmic}~(f)], due to the $dI_{cdw}/dV_z$ peak in Fig.~\ref{fig:non-Ohmic}~(d).

{\color{blue}\emph{Discussions and perspectives.}}
At higher magnetic fields, signatures of fractional quantum Hall effect have been reported \cite{Tang19nat,Galeski20arxiv}, which is a promising topic.
At lower magnetic fields ($B\in [0.6,1]$ T), the experiment also shows some plateau-like behaviors in the Hall resistivity~\cite{Tang19nat}, implying a simultaneous CDW phase of multiple bands.
The CDW mechanism of 3D quantum Hall effect could be realized also in layered structures HfTe$_5$, TaS$_2$, NbSe$_3$, etc. In Type-II Weyl semimetals \cite{Trescher17prbrc}, the overtilted pockets may lead to a cascade of CDW and even multiple 3D Hall plateaus for weak interactions.

\begin{acknowledgments}We thank helpful discussions with Liyuan Zhang, Kun Jiang, Fanqi Yuan, Changle Liu, Lianyi He, Peng-Lu Zhao, and Rui Chen. This work was supported by the National Natural Science Foundation of China (Grants No. 11534001, No. 11974249, and No. 11925402), the Strategic Priority Research Program of Chinese Academy of Sciences (Grant No. XDB28000000), Guangdong province (Grants No. 2016ZT06D348, No. 2017ZT07C062, No. 2020KCXTD001),  the National Key R \& D Program (Grant No. 2016YFA0301700), the Natural Science Foundation of Shanghai (Grant No. 19ZR1437300), Shenzhen High-level Special Fund (Grants No. G02206304 and No. G02206404), and the Science, Technology and Innovation Commission of Shenzhen Municipality (No. ZDSYS20190902092905285, No. ZDSYS20170303165926217, No. JCYJ20170412152620376, and No. KYTDPT20181011104202253). F.Q. acknowledges support from the project funded by the China Postdoctoral Science Foundation (Grants No. 2019M662150 and No. 2020T130635) and SUSTech Presidential Postdoctoral Fellowship. The numerical calculations were supported by Center for Computational Science and Engineering of Southern University of Science and Technology. The reuse of Fig.~\ref{fig:DeltaBT0}~(a) and Fig.~\ref{fig:non-Ohmic}~(h) has been approved by Springer Nature under Licence Number 4782250194315.
\end{acknowledgments}

%
\end{document}